\documentclass[prc,onecolumn,showpacs,amsmath,amssymb]{revtex4}
\usepackage{graphics,epsfig}
\makeatletter

\begin{document}
\title{New determination of the spectroscopic factor of $^7$Be ground state and the $^6$Li($p,\gamma$)$^7$Be astrophysical S(E) factors}

\author{LI Zhi-Hong}\email{zhli@ciae.ac.cn}
\author{SU Jun}
\author{LI Yun-Ju}
\author{WANG You-Bao}
\author{YAN Sheng-Quan}
\author{GUO Bing}
\author{NAN Ding}
\author{LIU Wei-Ping}
\affiliation{$^1$ China Institute of Atomic Energy, PO Box 275(46), Beijing 102413, China}

\author{LI Er-Tao}
\affiliation{$^2$ College of Physics and Energy, Shenzhen University, Shenzhen 518060, China}

\begin{abstract}
The `lithium problem' in Big Bang Nucleosynthesis (BBN) has recently focused on the reactions involving $^7$Be. The $^6$Li($p, \gamma$)$^7$Be reaction can provide us not only the information for destroying $^6$Li but also the information for producing $^7$Be. In the present work, the proton spectroscopic factor in $^7$Be was extracted to be 0.70 $\pm$ 0.17 from the angular distribution of $^7$Be($d$, $^3$He)$^6$Li at $E_\mathrm{c.m.}$ = 6.7 MeV. The value was then used to compute the direct component of the astrophysical $^6$Li($p, \gamma$)$^7$Be$_\mathrm{g.s.}$ S(E) factors and determine the resonance parameters from the total S(E) factors.

\end{abstract}
\pacs{26.35.+c, 21.10.Jx, 25.40.Lw}

\maketitle

\section{Introduction}
According to the standard Big Bang model, the universe starts from a singularity of extremely high temperature and density. The primordial nucleosynthesis takes place between 10 seconds and 20 minutes right after Big Bang. Big Bang Nucleosynthesis (BBN)~\cite{Fields2004} is the starting point of the elements, which can tell us not only the evolution of the elements but also the thermal history of the early universe. In standard theory of BBN (SBBN), the abundances of $^2$H, $^3$He, $^4$He and $^7$Li depend on only one cosmological parameter, the baryon-to-photon ratio, which can be constrained with high accuracy measurement of the Cosmic Microwave Background (CMB). Using the data from the precision observations of the CMB radiation with the Wilkinson Microwave Anisotropy Probe (WMAP)~\cite{Pryke2002, Komatsu2011}, the BBN predictions for the primordial abundances of $^2$H and $^4$He are in good agreement with the observations. However, for $^7$Li, there is a significant discrepancy between BBN predictions and the abundance derived from metal poor halo stars~\cite{Cyburt2008}.

The results from SBBN network calculations~\cite{Zhli2011} showed that the primordial $^7$Li were mainly produced from $^7$Be via the electron capture decay. Such been the case, the `lithium problem' in BBN should be focused on the reactions involving $^7$Be. The $^3$He($\alpha,\gamma$)$^7$Be reaction is the leading process to produce $^7$Be, which has been studied with great efforts~\cite{Holmgren1959, Parker1963, Nagatani1969, Krawinkel1982, Robertson1983, Volk1983, Alexander1984, Osborne1984, Hilgemeier1988}. As a supplementary reaction to produce $^7$Be, the $^6$Li($p, \gamma$)$^7$Be reaction, which is crucial for the consumption of $^6$Li and the formation of $^7$Be, has also attracted wide attention in the past years~\cite{Bashkin1955, Warren1956, Sweeney1969, Switkowski1979, Barker1980, Tingwell1987, Cecil1992, Angulo1999, Arai2002, Prior2004, Camargo2008, Huang2010, Dubovichenko2011, Nesterov2011, Xu2013, Hejj2013, Dong2017}. It is commonly believed that the direct capture dominates the $^6$Li($p, \gamma$)$^7$Be reaction at low energy. However, He et al.~\cite{Hejj2013} found a broad resonance in the astrophysical interesting energy region in 2013, which will change the evaluation for the contribution of this reaction to the big bang nucleosynthesis and the $^7$Be($p, \gamma$)$^8$B solar neutrino reaction. The reproduced astrophysical S(E) factors using R-matrix method can not well describe the experimental data, and further study will help us to understand the properties of this low energy resonance.

In the present article, we will reanalyze the angular distribution of $^7$Be($d$, $^3$He)$^6$Li measured in inverse kinematics with the secondary $^7$Be beam, which was described detail in our previous work~\cite{Liet2018}. The proton spectroscopic factor in $^7$Be ground state is extracted with the distorted wave Born approximation (DWBA) analysis and then used to compute the direct capture component of the astrophysical $^6$Li($p, \gamma$)$^7$Be S(E) factors. The S(E) factors measured by He et al.~\cite{Hejj2013} and Switkowski et al.~\cite{Switkowski1979} were then reanalyzed with our current experimental results.

\section{Extracting the proton spectroscopic factor in $^7$Be ground state}

The differential cross sections of $^7$Be($d$, $^3$He)$^6$Li at $E_\mathrm{c.m.}$ = 6.7 MeV were measured using the second beam facility~\cite{Wpliu2003} of HI-13 tandem accelerator in Beijing. The experimental setup is similar to previous~\cite{Wpliu1996, Zhli2005, Zhli2006} experiments, and the detailed description can be found in Ref.~\cite{Liet2018}. Here, we focus on the extracting of the $^7$Be proton spectroscopic factor from the angular distribution of $^7$Be($d$, $^3$He)$^6$Li.

The spins and parities of $^{6}$Li and $^{7}$Be (ground state) are
$1^{+}$ and $3/2^{-}$, respectively. The cross section of the
$^6$Li($p, \gamma$)$^7$Be reaction is comprised of the $s$-wave
proton transition to $1p3/2$ and $1p1/2$ orbit in $^{7}$Be ground state. The relations between the experimental differential cross section and the one from DWBA calculation can be expressed as
\begin{equation}\label{eq1}
\sigma_\mathrm{exp} = S_\mathrm{^3He}[S^\mathrm{^7Be}_{p3/2}\sigma_{p3/2}(\theta)+S^\mathrm{^7Be}_{p1/2}\sigma_{p1/2}(\theta)],
\end{equation}
where $\sigma_{exp}$ and $\sigma_{lj}(\theta)$
denote the measured and theoretical differential cross sections
respectively. $S^\mathrm{^{7}Be}_{lj}$ and $S_{^3He}$ stand for the nuclear
spectroscopic factors for the $^{7}$Be $\rightarrow$ $^{6}$Li + $p$ and $^3$He
$\rightarrow$ $d$ + $p$ virtual decays. By knowing the value of $S_\mathrm{^3He}$ , the
$S^\mathrm{^{7}Be}_{p3/2}$ and $S^\mathrm{^{7}Be}_{p1/2}$ can then be extracted by normalizing the theoretical differential cross sections to the experimental data with Eq.~(\ref{eq1}).

\begin{figure}
\includegraphics[width=8.0 cm]{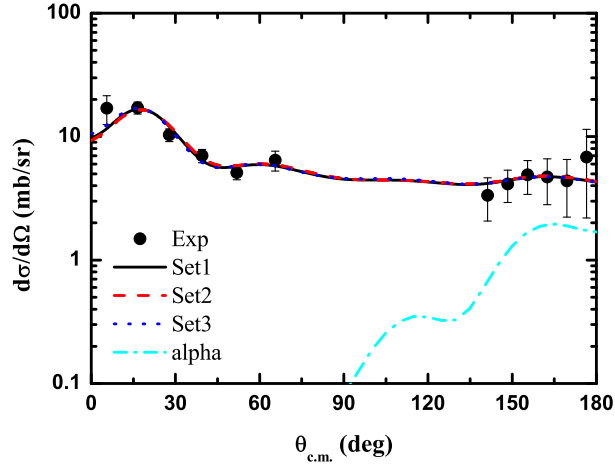}
\caption{\label{fig1} (Color online) The experimental and calculated angular distributions of $^7$Be($d$, $^3$He)$^6$Li at $E_\mathrm{c.m.}$ = 6.7 MeV. The solid circles represent the experimental data from the present work. The curves with different colors are the calculation results with three sets of optical potential parameters and the alpha transfer channel.}
\end{figure}

The DWBA calculation code TWOFNR \cite{Twofnr5} is adopted to obtain the theoretical differential cross sections. The spectroscopic factor of $^3$He $\rightarrow$ $d$ + $p$ has already been embedded in the code. The peripheral amplitudes of the reaction make a dominant contribution to the differential cross sections in the forward peak region. Therefore, the differential cross sections at the forward angles are used to extract the spectroscopic factor of $^7$Be. The compound nucleus contribution, which has little impact on the spectroscopic factor, can be considered to be isotropic in the present DWBA calculations. The optical potential parameters for both entrance and exit channels are listed in Table~\ref{tab1}. These parameters are taken from Ref.~\cite{Perey1976} and Ref.~\cite{Daehnick1980}. For the convenience of the calculations, these potential parameters have been put in TWOFNR code. With the theoretical ratio of $S^\mathrm{^7Be}_{p3/2}$ and $S^\mathrm{^7Be}_{p1/2}$, the spectroscopic factors in ground state of $^7$Be are deduced to be $S^\mathrm{^7Be}_{p3/2}$ = 0.47 $\pm$ 0.10, 0.41 $\pm$ 0.09, 0.38 $\pm$ 0.09 and $S^\mathrm{^7Be}_{p1/2}$ = 0.31 $\pm$ 0.07, 0.28 $\pm$ 0.06, 0.25 $\pm$ 0.06 by the three sets of optical potential parameters. The average values are 0.42 $\pm$ 0.10 and 0.28 $\pm$ 0.07, and the errors are mainly resulted from the uncertainties of optical potential parameters (10\%) and the statistics (22\%). Therefore the total proton spectroscopic factor $S^\mathrm{^7Be}_{tot}$ = $S^\mathrm{^7Be}_{p3/2}$ + $S^\mathrm{^7Be}_{p1/2}$ can be deduced to be 0.70 $\pm$ 0.17, and the corresponding ANC is 1.84 $\pm$ 0.45 fm$^{-1/2}$. The alpha transfer reaction channel of $^7$Be($d$,$^6$Li)$^3$He is also taken into account in the present calculation. It has little effect on the extracted spectroscopic factor because the cross sections of alpha transfer reaction are about two orders of magnitude smaller than the one nucleon transfer reaction at forward angles, but it can reproduce the differential cross section at the backward angles well. The normalized angular distributions with these optical potential parameters are presented in Fig.~\ref{fig1}. The present spectroscopic factors are in good agreement with the shell-model values~\cite{Cohen1967}, the Green's function Monte Carlo calculational values~\cite{Brida2011}, the value extracted from the $^6$Li($^3$He,$d$)$^7$Be angular distributions by Burtebayev et al.~\cite{Burtebayev2013}, and our previous results with the $^7$Li($^6$Li, $^7$Li)$^6$Li elastic transfer reaction~\cite{Jsu2010}.

\begin{table}
\caption{ \label{tab1} The optical potential parameters used in the DWBA calculations, the Coulomb radius parameter $r_C$ = 1.3 is adopted for all channels. V and W are the depths in MeV, and r and $a$ are the radius and diffuseness in fm.}
\footnotesize
\begin{tabular*}{160mm}{c@{\extracolsep{\fill}}ccccccccccccc}
\hline
Channel & V & $r_V$ & $a_V$ & W & $r_W$ & $a_W$ & $W_s$ & $r_s$ & $a_s$ & $V_{SO}$ & $r_{SO}$ & $a_{SO}$ & Ref. \\
\hline\hline
$d + ^7$Be      & 95.7  & 1.05 & 0.86 &      &      &      & 59.6 & 1.43 & 0.55 & 3.5 & 0.75 & 0.50 & \cite{Perey1976}\\
$d + ^7$Be      & 83.9  & 1.15 & 0.81 &      &      &      & 16.5 & 1.34 & 0.68 &     &      &      & \cite{Perey1976}\\
$d + ^7$Be      & 88.1  & 1.17 & 0.72 & 0.09 & 1.33 & 0.67 & 12.3 & 1.33 & 0.67 & 3.5 & 1.07 & 0.66 & \cite{Daehnick1980}\\
$^3$He + $^6$Li & 150.2 & 1.20 & 0.72 & 38.4 & 1.40 & 0.88 &      &      &      & 2.5 & 1.20 & 0.72 & \cite{Perey1976}\\
\hline
\end{tabular*}
\end{table}

\section{Astrophysical $^6$Li($p, \gamma$)$^7$Be S(E) factor}

According to the experimental data in Ref.~\cite{Hejj2013}, there are two main processes in the proton radioactive capture reactions at the stellar energies, e.g. the direct capture and the resonant capture processes. The direct capture of the $^6$Li($p,\gamma$)$^7$Be reaction is dominated by the $E1$ transition of the proton from incoming $s$ wave to bound $p$ state. The cross section can be computed using the traditional direct capture model~\cite{Rolfs73,Christy61,Zhli2006}
\begin{eqnarray}\label{eq2}
\sigma_\mathrm{dc}&=&\frac{16\pi}{9}\left(\frac{E_{\gamma}}{\hbar
c}\right)^3\frac{e^{2}_{\mathrm{eff}}}{k^2}\frac{1}{\hbar
v}\frac{(2I_{f}+1)}{(2I_{1}+1)(2I_{2}+1)}S_{lj}\nonumber\\&
&\times\left|\int\limits^\infty_0
r^{2}w_{l_i}(kr)u_{l_f}(r)dr\right|^2,
\end{eqnarray}
where $E_\gamma$ is the $\gamma$-ray energy. $e_{\mathrm{eff}}=eN/A$ stands for the proton effective charge for
the E1 transition for a target nucleus ($A$, $Z$). $v$ is the relative velocity between proton and $^6$Li. $k$ denotes the wave number of proton. $I_1$, $I_2$ and $I_f$ are the spins of proton, $^6$Li and $^7$Be, respectively. $w_{l_i}(kr)$ refers to the distorted radial wave function for the continuum, and $u_{l_f}(r)$ the radial wave function of the bound state of $^7$Be. The astrophysical S(E) factor is a rescaled variant of the cross section that accounts for the Coulomb repulsion between the charged reactants. It is defined as
\begin{equation}
\label{eq3}%
S(E)=E\sigma(E)\exp(E_{G}/E)^{1/2},
\end{equation}
where the Gamow energy $E_{G}=0.978Z^{2}_{1}Z^{2}_{2}\mu$ MeV, $\mu$ is the reduced mass of the system.

\begin{figure}
\includegraphics[width=8.2 cm]{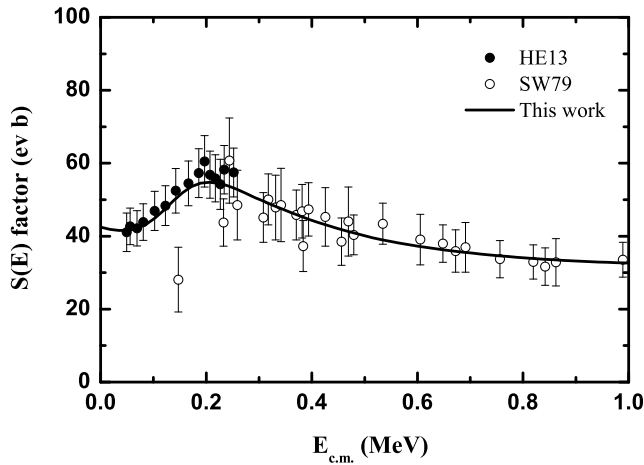}
\caption{\label{fig2} The astrophysical S(E) factors of $^6$Li($p, \gamma$)$^7$Be$_\mathrm{g.s.}$. The solid curve is the best fitting result in the present work. The data with solid circles represent the experimental S(E) factors by He et al.~\cite{Hejj2013}, those with open circles are determined by Switkowski et al.~\cite{Switkowski1979}. All the experimental data are multiplied by the branching ratio to the ground state. }
\end{figure}

Using the spectroscopic factor ($S_{lj}$ = 0.70 $\pm$ 0.17) deduced from the $^{7}$Be($d, ^3$He)$^{6}$Li
transfer reaction, the cross section and the astrophysical S(E) factor for the direct capture process of the $^6$Li($p, \gamma$)$^7$Be reaction can then be calculated by Eq.~(\ref{eq2}) and Eq.~(\ref{eq3}). The code FRESCO~\cite{FRESCO} was adopted in the present calculations. The wave functions for both bound and continuum states were computed by solving the Schr\"{o}dinger equation using a Woods-Saxon form potential with the standard geometrical parameters ($r$ = 1.25, $a$ = 0.65). The potential depth for the bound state is adjusted to reproduce the binding energy. For the continuum potential, the depth can be fixed by scaling the direct component of $^6$Li($p, \gamma$)$^7$Be, which can be determined by the edge of the broad resonant peak. The influence of the imaginary potential is very small comparing to the real potential and thus can be neglected in the calculation. The direct component ($S_\mathrm{dc}$) of the astrophysical S(E) factors deduced with the above process are presented by the plot in blue color in Fig.~\ref{fig2} .

In order to explain the total S(E) factors of $^6$Li($p, \gamma$)$^7$Be, the contribution of the broad resonance at low-energy region is indispensable. The cross section of the one-level resonance capture can be expressed by Breit-Wigner formula
\begin{equation}
\label{eq4}
\sigma_{rc}=\frac{\pi}{k^2}\frac{\omega\Gamma_p(E)\Gamma_\gamma(E)}{(E-E_r)^2+\Gamma_t^2(E)/4},
\end{equation}
where $E_r$ is the resonant energy. $\omega$ represents the production of the statistical factor, which can be calculated with the spin parameters by the expression
\begin{equation}
\label{eq5}
\omega=\frac{2J_f+1}{(2J_p+1)(2J_t+1)},
\end{equation}
$\Gamma_{p}(E)$, $\Gamma_{\gamma}(E)$ and $\Gamma_t(E)$ are the observable partial width of the resonance in the channel $^6$Li + $p$, the observable radiative width for the decay of the given resonance into the ground state of $^7$Be, and the total width, respectively. The three widths are all energy dependent, their relations to the experimental partial and radiative widths can be found in Ref.~\cite{Zhli2006}. The astrophysical S(E) factor of the resonant component can also be computed with Eq.~(\ref{eq3}).

The S(E) factors of $^6$Li($p, \gamma$)$^7$Be$_\mathrm{g.s.}$ have been calculated by using a simple direct-resonant interference model~\cite{Rolfs73}. The analysis can produce the best fit with many unphysical parameter sets, thus more known parameters are need to give meaningful results. It is very difficult to measure the low energy resonance, especially for the hundreds keV resonance in $^7$Be. Bouchez et al.~\cite{Bouchez1960} had introduced two low-energy states in $^7$Be from the determination of the $^6$Li($p$, $^3$He)$^4$He angular distributions. Neither of these states is observed in $p$-$^6$Li scattering and $^3$He¨C$^4$He scattering~\cite{Brown1968}. Mani and Dix~\cite{Mani1968} reported a level of unknown spin and parity in $^7$Li by $p$ + $^7$Li scattering, the corresponding level in the mirror nucleus $^7$Be would lie near the binding energy of $^6$Li + $p$. In 2013, He et al.~\cite{Hejj2013} found a level at 195 keV from the proton capture reaction. The proposed resonant parameters can't reproduce the $^6$Li($p, \gamma$)$^7$Be S(E) factor very well. In such case, we instead do the analysis with R-matrix code AZURE~\cite{Azuma2010}, which is designed to model low-energy nuclear reactions involving charged particles, $\gamma$-rays, and neutrons. The code allows for more strict constraints on the fitted parameters than the simple direct-resonant interference model.

 With the direct capture S(E) factors obtained from the present work and the S(E) factors measured by He et al.~\cite{Hejj2013} and Switkowski et al.~\cite{Switkowski1979}, the resonance parameters can then be obtained by fitting the $^6$Li($p, \gamma$)$^7$Be$_\mathrm{g.s.}$ S(E) factors. To do this, we think about two possible cases, namely the resonance dominated by the proton width or by the alpha width. In the analysis, the channel radius is fixed to be 3.4 fm. For the first case, we get the same result as He et al.~\cite{Hejj2013}. For the second case, our best fitting results are ($E_{r}$ = 145 keV, $\Gamma_{p}(E_{r})$ = 10.2 eV, $\Gamma_{\gamma}(E_{r})$ = 7.6 eV and $\Gamma_{\alpha}(E_{r})$ = 232 keV ). The testing for goodness of fit, $\chi^2_{min}$, is 15.4. One can see from Fig.~\ref{fig2} that the resonant parameters extracted from the present work can reproduce the total S(E) factors of $^6$Li($p, \gamma$)$^7$Be very well.

\section{Summary and conclusion}

In summary, we have extracted the proton spectroscopic factor of $^7$Be ground state from the angular distribution of $^{7}$Be($d, ^3$He)$^{6}$Li, and deduced the direct capture components of the $^6$Li($p, \gamma$)$^7$Be reaction. The data have been used to fit the experimental total S(E) factors~\cite{Hejj2013} and obtained the resonance parameters of the 3/2$^+$ broad resonance level in $^7$Be.

The existence of the broad resonance in the 1$p$ shell nuclei had been observed in the $^{11}$C($p, \gamma$)$^{12}$N~\cite{Tang2003}, $^{12}$C($p, \gamma$)$^{13}$N~\cite{Rolfs73}, $^{12}$N($p, \gamma$)$^{13}$O~\cite{Bguo2013}, and $^{13}$N($p, \gamma$)$^{14}$O~\cite{Zhli2006} reactions. Such situation may be common for 1$p$ shell nuclei, and therefore affect the stellar nucleosynthesis for the light nuclei.

\section{The Acknowledgements}
Supported by National Natural Science Foundation of China (11375269, 11490563 and 11505117), Natural Science Foundation of Guangdong Province under Grant No. 2015A030310012 and National Basic Research Program of China (2013CB834406) and National Key Research and Development Program of China (2016YFA0400502).


\begin{thebibliography}{90}
\vspace{3mm}
\bibitem{Fields2004}  B. Fields, K. Olive, Nucl. Phys. A {\bf 777}£º208¨C225 (2006)
\bibitem{Pryke2002} C. Pryke et al., Astrophys. J. {\bf 568}: 46 (2002)
\bibitem{Komatsu2011} E. Komatsu et al. [WMAP Collaboration], Astrophys. J. Suppl. {\bf 192}: 18 (2011)
\bibitem{Cyburt2008} R. Cyburt, B. Fields and K. Olive, J. Cosm. Astro. Phys. {\bf 11}: 012 (2008)
\bibitem{Zhli2011} Z. H. Li et al., Sci. China Phys. Mech. Astron. {\bf 54}: s67 (2011)
\bibitem{Ryan1999} S. G. Ryan et al., Astrophys. J. {\bf 523}: 654 (1999)
\bibitem{Holmgren1959}H. D. Holmgren and R. L. Johnston, Phys. Rev. {\bf 113}: 1556 (1959)
\bibitem{Parker1963} P. D. Parker and R. W. Kavanagh, Phys. Rev. {\bf 131}: 2578 (1963)
\bibitem{Nagatani1969} K. Nagatani, M. R. Dwarakanath, and D. Ashery, Nuclear Physics A {\bf 128}: 325 (1969)
\bibitem{Krawinkel1982} H. Kr\"{a}winkel  et al., Zeitschrift f\"{u}r Physik A Hadrons and Nuclei {\bf 304}: 307 (1982)
\bibitem{Robertson1983} R. G. H. Robertson  et al., Phys. Rev. C {\bf 27}: 11 (1983)
\bibitem{Volk1983} H. Volk, H. Kr\"{a}winkel, R. Santo, and L. Wallek, Zeitschrift fur Physik {\bf 310}: 91 (1983)
\bibitem{Alexander1984} T. K. Alexander et al., Nuclear Physics A {\bf 427}: 526 (1984)
\bibitem{Osborne1984} J. L. Osborne et al., Nuclear Physics A {\bf 419}: 115 (1984)
\bibitem{Hilgemeier1988} M. Hilgemeier et al., Zeitschrift f\"{u}r Physik A Hadrons and Nuclei {\bf 329}: 243 (1988)
\bibitem{Bashkin1955} S. Bashkin and R. R. Carlson, Phys . Rev . {\bf 97}: 1245 (1955)
\bibitem{Warren1956} J. B. Warren, T. K. Alexander and G. B. Chadwick, Phys. Rev. {\bf 101}: 242 (1956)
\bibitem{Sweeney1969} W. E. Sweeney, Bull . Am. Phys . Soc. {\bf 14}: 487 (1969)
\bibitem{Switkowski1979} Z. E. Switkowski et al, Nucl. Phys. A {\bf 331}: 50-60 (1979)
\bibitem{Barker1980} F. C. Barker, Aust. J. Phys. {\bf 33}: 159 (1980)
\bibitem{Tingwell1987} C.I.W. Tingwell, J.D. King, and D.G. Sargood, Aust. J. Phys. {\bf 40}: 319-328 (1987)
\bibitem{Cecil1992} F. E. Cecil, et al., Nucl. Phys. A {\bf 539}: 75 (1992)
\bibitem{Angulo1999} C. Angulo, et al., Nucl. Phys. A {\bf 656}: 3 (1999)
\bibitem{Arai2002} K. Arai, D. Baye, P. Descouvemont, Nucl. Phys. A {\bf 699}: 963¨C975 (2002)
\bibitem{Prior2004} R. M. Prior et al., Phys. Rev. C {\bf 70}: 055801 (2004)
\bibitem{Camargo2008} O. Camargo, et al., Phys. Rev. C {\bf 78}: 034605 (2008)
\bibitem{Huang2010} J. T. Huang, C. A. Bertulani, and V. Guimar\~{a}es, At. Data Nucl. Data Tables {\bf 96}: 824 (2010)
\bibitem{Dubovichenko2011} S. B. Dubovichenko et al., Phys. At. Nucl. {\bf 74}: 984 (2011)
\bibitem{Nesterov2011} A. V. Nesterov, V. S. Vasilevsky, and T. P. Kovalenko, Ukr. J. Phys. {\bf 56}: 645 (2011)
\bibitem{Xu2013} Y. Xu et al., Nucl. Phys. A {\bf 918}: 61 (2013)
\bibitem{Hejj2013} J. J. He et al, Phys. Lett. B {\bf 725}: 287-291 (2013)
\bibitem{Dong2017} G. X. Dong et al, Journal of Physics G: Nuclear and Particle Physics, {\bf 44}: 045201 (2017)
\bibitem{Liet2018} E. T. Li et al, Chin. Phys. C {\bf 42}: 044001 (2018)
\bibitem{Wpliu2003} W. P. Liu et al, Nucl. Instrum. Methods B {\bf 204}: 62 (2003)
\bibitem{Wpliu1996} W. P. Liu et al, Phys. Rev. Lett. {\bf 77}: 661 (1996)
\bibitem{Zhli2005} Z. H. Li et al, Phys. Rev. C {\bf 73}: 052801 (2005)
\bibitem{Zhli2006} Z. H. Li et al, Phys. Rev. C {\bf 74}: 035801 (2006)
\bibitem{Twofnr5} M. Igarashi et al., Computer Program TWOFNR (Surrey University version).
\bibitem{Perey1976} C. M. Perey and F. G. Perey, Atomic Data and Nuclear Data Tables {\bf 17}: 1 (1976)
\bibitem{Daehnick1980} W. W. Daehnick, J. D. Childs, and Z. Vrcelj, Phys. Rev. C {\bf 21}: 2253 (1980)
\bibitem{Cohen1967} S. Cohen and D. Kurath, Nucl. Phys. A {\bf 101}: 1 (1967)
\bibitem{Brida2011} I. Brida, S. C. Pieper and R. B. Wiringa, Phys. Rev. C {\bf 84}: 024319 (2011)
\bibitem{Burtebayev2013} N. Burtebayev et al., Nucl. Phys. A {\bf 909}: 20 (2013)
\bibitem{Jsu2010} J. Su et al., Chin. Phys. Lett. {\bf 27}: 052101 (2010)
\bibitem{Christy61} R. F. Christy and I. Duck, Nucl. Phys. A {\bf 24}: 89 (1961)
\bibitem{Rolfs73} C. Rolfs, Nucl. Phys. A {\bf 217}: 29 (1973)
\bibitem{FRESCO} I. Thompson, Computer Physics Reports {\bf 7}: 167-212 (1988)
\bibitem{Bouchez1960} R. Bouchez et al., J. de Phys. {\bf 21}: 346 (1960)
\bibitem{Brown1968} L. Brown and C. Petitjean, Nucl. Phys. A {\bf 117}: 343-353 (1968) and the references therein.
\bibitem{Mani1968} G. Mani and A. Dix, Nucl. Phys. A {\bf 106}: 251-260 (1968)
\bibitem{Azuma2010} R. E. Azuma et al., Phys. Rev. C {\bf 81}: 045805 (2013)
\bibitem{Tang2003} X. Tang et al., Phys. Rev. C {\bf 67}: 015804 (2003)
\bibitem{Bguo2013} B. Guo et al., Phys. Rev. C {\bf 87}: 015803 (2013)

\end{thebibliography}
\end{document}